\documentstyle[galley, rotate]{mn}
\onecolumn
\input{epsf}

\title[Generalised Cumulant Correlators]
{Generalised Cumulant Correlators and Hierarchical Clustering}

\author[D. Munshi et al.]{Dipak Munshi$^{1,4}$,
Adrian L. Melott$^2$ and Peter Coles$^{1,3}$\\ $^1$Queen Mary and
Westfield College, London E1 4NS, United Kingdom \\ $^2$Department
of Physics and Astronomy, University of Kansas, Lawrence, Kansas
66045, U.S.A., \\ $^3$School of Physics \& Astronomy, University
of Nottingham, University Park, Nottingham, NG7 2RD, United
Kingdom\\ $^4$International School for Advanced Studies (SISSA),
Via Beirut 2-4, I-34013 Trieste, Italy}

\pagerange{000,000}
\begin{document}

\maketitle

\begin{abstract}
The cumulant correlators, $C_{pq}$, are statistical quantities
that generalise the better-known $S_p$ parameters; the former are
obtained from the two-point probability distribution function of
the density fluctuations while the latter describe only the
one-point distribution. If galaxy clustering develops from
Gaussian initial fluctuations and a small-angle approximation is
adopted, standard perturbative methods suggest a particular
hierarchical relationship of the $C_{pq}$ for projected clustering
data, such as the APM survey. We establish the usefulness of the
two-point cumulants for describing hierarchical clustering by
comparing such calculations against available measurements from
projected catalogs, finding very good agreement. We extend the
idea of cumulant correlators to multi-point {\em generalised}
cumulant correlators (related to the higher-order correlation
functions). We extend previous studies in the highly nonlinear
regime to express the generalised cumulant correlators in terms of
the underlying ``tree amplitudes'' of hierarchical scaling models.
Such considerations lead to a technique for determining these
hierarchical amplitudes, to arbitrary order, from galaxy catalogs
and numerical simulations. Knowledge of these amplitudes yields
important clues about the phenomenology of gravitational
clustering. For instance, we show that three-point cumulant
correlator can be used to separate the tree amplitudes up to sixth
order. We also combine the particular hierarchical {\em ansatz} of
Bernardeau \& Schaeffer (1992) with extended and hyperextended
perturbation theory to infer values of the tree amplitudes in the
highly nonlinear regime.
\end{abstract}


\section{Introduction}
%\section{hierarchical Models of Clustering at nonlinear regime}

The Newtonian gravitational force is scale-free. This leads one
directly to the suspicion that, at least in the strong clustering
limit where memory of any specific length scale set by the initial
conditions has been obliterated, the complex pattern of
gravity-driven clustering should  display relatively simple
scaling properties. One way such scaling should manifest itself is
in the behaviour of the hierarchy of $N$-point correlation
functions (e.g. Peebles 1980). This has led to the formulation of
the so-called hierarchical {\em ansatz} for gravitational
clustering which, in its most general form, can be written ,
\begin{equation}
\xi_N( \lambda {\bf r_1}, \dots  \lambda {\bf r_N} ) =
\lambda^{-\gamma(N-1)} \xi_N( {\bf r_1}, \dots, {\bf r_N} )
\label{hierarchical}
\end{equation}
(e.g. Balian \& Schaeffer 1989), where $\xi_N$ is the $N$-point
correlation function for particles located at positions ${\bf r_1}
\ldots {\bf r_N}$, and $\gamma$ is the negative slope of two-point
correlation function $\xi({\bf r_i}, {\bf r_j})$. A more
restrictive, but still quite general, algebraic form is often used
in the literature, in which the correlation function is
constructed from linear superposition of all possible topologies
of ``tree'' diagrams connecting $N$ points with $N-1$ edges. This
model permits the association of a number with each distinct tree
topology (the tree amplitude) with the amplitudes themselves left
arbitrary (Balian \& Schaeffer 1989). In other words,
\begin{equation}
\xi_N( {\bf r_1}, \dots {\bf r_N} ) = \sum_{\alpha, \rm N-trees}
Q_{N,\alpha} \sum_{\rm labellings} \prod_{\rm edges}^{(N-1)}
\xi({\bf r_i}, {\bf r_j}),
\end{equation}
where the last term is the two-point correlation function for each
pair of particles $(i,j)$. It is important to note that this model
is not the only possible way to formulate hierarchical clustering.
In general, the  amplitudes $Q_{N,\alpha}$ could depend on the
geometry of a specific configuration of $N$ points, and not just
its topology as is assumed here. However, there are some
indications, at least at the level of three points, that the
amplitudes do become independent of configuration in the highly
non-linear regime (Scoccimarro et al. 1998).
There are some
indications that they become independent of level of non-linearity
too at highly non-linear regime (Munshi et al. 1999; see Colombi et al.
1996 for a different view-point).

Many different models of hierarchical scaling, largely consisting
of particular choices for the $Q_{N,\alpha}$, can be found in the
literature. In principle, it should be possible to derive these
amplitudes from the underlying gravitational physics encoded in
the hierarchy of BBGKY equations that describe the dynamical
origin of gravity-driven clustering  (Davis \& Peebles 1977; Fry
1984; Hamilton 1988). However, no general solution to these
equations exists, so these various approximate models  are
employed in order to get some insight into the physics of
gravitational clustering in highly nonlinear regime and to aid
interpretation of empirical galaxy clustering data.

Despite the increasing availability of large samples of galaxy
redshifts, the most precise statistical analyses of galaxy
clustering are still performed using projected (angular) samples.
The extremely large size of projected galaxy surveys, together
with the relatively easy-to-handle effect of projection, makes
such samples ideal for the extraction of descriptors such as
higher-order cumulants. The APM survey (Maddox et al. 1990;
Maddox, Sutherland \& Efstathiou 1990; Maddox, Efstathiou \&
Sutherland 1996), which contains more than 1.3 million objects, is
at present the largest angular catalogue available. Such is the
amount of useful information in this sample that Gaztanaga (1994)
was able to extract estimates of the so-called $S_p$ parameters
(see below), for $p\leq 9$ from it. Such an analysis is at present
beyond the scope of galaxy redshift surveys (Bouchet et al. 1993),
at least until the Sloan Digital Sky Survey and Anglo-Australian
2DF survey are completed. Projected catalogs are also free from
redshift-space distortions, a fact which can be exploited to yield
accurate estimates of real-space clustering phenomena in a
relatively unambiguous way. The obvious difficulty associated with
projected catalogs is related to the fact that any projected
angular scale corresponds to a superposition of physical length
scales which have which represent different levels of
nonlinearity. Limber (1954) pioneered the study of relating
angular correlation functions to their real-space counterparts.
These inversion techniques and their small angle approximations
are now known to be very efficient for the extraction of
information from projected catalogs (e.g. Peebles 1980; Baugh \&
Efstathiou 1993).

Many statistical studies of large-scale structure focus (in one
way or another) on the  determination of one-point cumulants for
cells of a particular size, which are essentially volume averages
of correlation functions over the cell. The second-order cumulant,
for example, is the variance of cell-counts, and which is the
average of the two-point correlation function over the cell. The
standard $S_p$ parameters (see below) are based on the cumulants
of the one-point probability distribution of density fluctuations,
$f_1(\delta)$, and can be expressed in terms of volume averages of
the higher-order correlation functions in an analogous fashion.
Here and throughout the paper $\delta({\bf x})$ denotes the
dimensionless density contrast, defined by
\begin{equation}
\delta({\bf x}) = \frac{\rho({\bf x})-\rho_0}{\rho_0},
\end{equation}
where $\rho_0$ is the mean density of matter.

Szapudi \& Szalay (1997) recently proposed a new class of
statistical measures, known as the cumulant correlators, which
generalises the set of $S_p$ parameters mentioned above. The
cumulant correlators $C_{pq}$ are analogous quantities obtained
from the two-point (bivariate) distribution $f_2(\delta_1,
\delta_2)$. They form a symmetric matrix, $C_{pq}$, and estimation
of these quantities from data is a very similar task to the
extraction of the standard two-point correlation function.
Furthermore, using the factorial moments of the discrete count
probability density function makes it possible to subtract the
effect of Poisson shot noise very efficiently. Szapudi \& Szalay
(1997) extracted estimates of these quantities for $p+q\leq 5$
from the APM galaxy survey.

Theoretical work on the cumulant correlators has been performed by
Bernardeau (1995), who obtained predictions for the $C_{pq}$ in
three-dimensional real space for clustering developing from
Gaussian initial conditions and in the limit of large spatial
separations. Interestingly, a relatively simple scaling hierarchy
is expected to develop under these conditions, identical to the
scaling of their one-point counterparts the $S_p$ parameters,
potentially furnishing a simple and powerful diagnostic of the
idea of gravitational instability and, perhaps, of the extent to
which galaxy bias affects clustering statistics.

In this paper we study the generalisation of the cumulant
correlators to multi-points  and for arbitrary order. These
statistics are constructed from the quantities $\langle
\delta^p({\bf x}_1) \delta^q({\bf x}_2) \dots \delta^s({\bf
x}_l)\rangle, $ but while the cumulant correlators have $l=2$ with
$p$ and $q$ arbitrary, the multipoint correlators have no
restriction on $l$ or any of the integers $p$, $q$, $\ldots s$. We
study the dependence of these quantities on the underlying
hierarchical tree amplitudes, with a view to testing the different
possible hierarchical models using these statistics. We also
present the results of a calculation of the quasilinear behaviour
of the cumulant correlators in projection, including the effects
of observational selection. In Section 2 we outline our main
theoretical results in highly nonlinear regime; Section 3 is
devoted to the computation of two-point cumulant correlators in
projected catalogues in quasi-linear regime. Theoretical results
are tested against measurements from the APM survey in Section 4.

\section{Generalized Cumulant Correlators in the Nonlinear Regime}

Such are the importance of hierarchical models in the
understanding of gravitational clustering that it is important to
learn which of choice of tree amplitudes in the model (2)
corresponds with reality. This is best tackled by attempting to
extract these amplitudes from numerical simulation data. However,
the direct determination of higher-order correlations is
complicated by their possible dependence on shape as well as
topology. Most numerical studies in this direction are therefore
have concentrated on an indirect determination these quantities,
using  one-point cumulants or two-point cumulant correlators. It
is known that cumulant correlators can be used to compute the
amplitudes associated with two different types of topologies
 -- the so-called ``snake'' and ``star'' topologies -- at fourth order (Szapudi
\& Szalay 1997). However, at higher orders than this the number of
different topologies rapidly increases and two-point cumulant correlators
are no longer sufficient to effect a unique separation of all
tree-amplitudes. The only possible way we can remedy the situation
is by moving to the generalised cumulant correlators mentioned
above. These are the natural generalisation of their two-point
counterparts (Munshi et al. 1998a) and, as we discuss elsewhere,
they are also related in a very interesting way to the statistics
of collapsed objects, in much the same way as the two-point
cumulant correlators are related to the natural bias associated
with treating collapsed objects rather than the general
distribution of mass (Bernardeau 1992; Munshi et al. 1998a).

The terminology in this field is a little confusing, so it is
worth beginning with a brief explanation of notation in order to
relate this work with the existing literature. For three points
there is only one possible tree, and the unique three--point
amplitude is generally denoted $Q$. As we mentioned in passing,
there are two possible topologies for connecting four points
(respectively called the ``snake'' and the ``star''). The
amplitudes of these tree configurations are usually denoted $R_a$
and $R_b$. Following in this vein we introduce, for the distinct
topologies at fifth order, the notations $S_a$, $S_c$ and $S_b$
 for the ``snake'', ``star'' and ``hybrid'' cases. At sixth order
 the  number of different topologies is six (Fry 1984) and one
can similarly use $T_a$, $T_b$, $T_c$, $T_d$, $T_e$ and $T_f$ to denote
the amplitudes associated with them.

The order $M$ of the multi-point cumulant correlator $\langle
\delta^p({\bf x}_1) \delta^q({\bf x}_2) \dots \delta^s({\bf
x}_l)\rangle$ is defined to be equal to $p+q+ \ldots + s$ and such
a quantity of order $M$ takes contributions from
$M$-point correlation function. Cumulant correlators of a particular order
will depend only
on hierarchical amplitudes associated with  topologies
contributing to the {\em same} order, even though the number of
spatial points ($l$) on which they depend may vary.

For example, the sixth-order two-point and three-point cumulant
correlators will both depend on tree amplitudes associated with
six-point correlation functions. For example, there are three
different but ``degenerate'' three-point cumulant correlators of
sixth order, a fact which stems from the number of different ways
the integer six can be decomposed in three integers, i.e.
$4+1+1=3+2+1=2+2+2=6$. All of these three-point cumulant
correlators in general can depend on all the six different
tree-amplitudes associated with sixth order correlation function.
Hence measurement of these three three-point cumulant correlators
of sixth order will provide us three equations connecting the six
tree amplitudes. The system of three equations with six variables
is consequently indeterminate, and can only be solved by including
three more equations connecting these amplitudes. But if we
compliment these equations with information provided  by the {\em
two-point} cumulant correlators of sixth order we get three more
equations, corresponding to the fact that $1+5=2+4=3+3=6$ and
quantities also depend on exactly the same tree amplitudes. The
resulting six equations in six variables will provide us with a
fully-determined set of equations which can be solved to yield the
amplitudes associated with all topologies of sixth order.

This example can be generalised. Counting the different number of
degenerate multi-point cumulant correlators, for a given order and
for a specific set of points, one can check that three-point
cumulant correlators can in principle be used to determine all
tree  amplitudes associated with correlation functions up to and
including sixth order. We have tabulated the different multi-point
cumulants of at a given order to show which can be used to
separate out the hierarchical amplitudes associated with
correlation functions through equation (2).

\subsection{Leading-order tree terms}

If we express two degenerate two-point cumulant correlators of
fourth order in the limit of {\em large separations} (or, in other
words, the variance evaluated at each point, related to
$\bar{\xi_2}$, is much bigger than the covariance between the
points, expressed as $\xi_{ab}$) we obtain following pair of
equations:
\begin{equation}
\langle  \delta^3 ({\bf x}_1)  \delta ({\bf x}_2)\rangle = (3R_b +
6R_a)\xi_{ab}\bar{ \xi_2}^2; \end{equation}
\begin{equation}
\langle \delta^2 ({\bf x}_1) \delta^2 ({\bf x}_2)\rangle = 4 R_b
\xi_{ab} \bar{ \xi_2}^2
\end{equation}
As explained earlier these two-point cumulant correlators depend
only on the two amplitudes associated with the four-point
correlation function, i.e. $R_a$ and $R_b$. They can solved
simultaneously to compute values of these quantities.

At fifth order there are also two degenerate two-point cumulant
correlators which depend on three tree-amplitudes:
\begin{equation}
\langle  \delta^4 ({\bf x}_1)  \delta ({\bf x}_2)\rangle = (4S_c +
36S_b + 24S_a)\xi_{ab}\bar{ \xi_2}^3 \end{equation}
\begin{equation}
\langle \delta^3 ({\bf x}_1) \delta^2 ({\bf x}_2)\rangle = (6 S_b
+ 12 S_a) \xi_{ab} \bar{ \xi_2}^3.
\end{equation}
To determine the amplitudes $S_a$, $S_b$ and $S_c$ the two-point
cumulant correlators would be insufficient, so we have to move to
the three-point quantities instead. These provide us another pair
of equations with same tree-amplitudes:
\begin{equation}
\langle  \delta^2 ({\bf x}_1)  \delta ({\bf x}_2)\delta^2 ({\bf
x}_3)\rangle = (4S_a \xi_{ab} \xi_{bc} + 4S_b \xi_{ab} \xi_{ac} +
4S_b \xi_{ac} \xi_{bc})\bar{ \xi_2}^2 \end{equation}
\begin{equation}
\langle \delta^3 ({\bf x}_1)  \delta ({\bf x}_2)\delta
({\bf x}_3)\rangle = ((6S_a + 3S_b)(\xi_{ab}\xi_{bc} +
\xi_{ac}\xi_{bc}) + (3S_c + 18S_b + 6S_a)\xi_{ab}\xi_{ac}) \bar{
\xi_2}^2
\end{equation}
Any three of the equations (6) to (9) can be used to determine the
tree-amplitudes $S_a, S_b$ and $S_c$. The other equation will be
consistent but redundant, and can be used as a check. Of course,
the one-point cumulant of same order always provide an additional
constraint at every order.

There are three degenerate two-point cumulant correlators of sixth
order and there are 6 distinct topologies:
\begin{equation}
\langle  \delta^4 ({\bf x}_1)  \delta^2 ({\bf x}_2)\rangle =
(48T_a + 48T_e + 24T_b + 8T_d)\xi_{ab}\bar{ \xi_2}^4
\end{equation}
\begin{equation}
\langle \delta^3 ({\bf x}_1)  \delta^3 ({\bf x}_2)\rangle = (36T_b
+ 36T_a + 9T_c) \xi_{ab} \bar{ \xi_2}^4
\end{equation}
\begin{equation}
\langle \delta^5 ({\bf x}_1) \delta ({\bf x}_2)\rangle =
(120T_a + 180T_b + 180T_e + 80 T_d + 60 T_c + 5T_f) \xi_{ab} \bar{
\xi_2}^4
\end{equation}
Combining these three equations with expressions for the
three-point cumulant correlators at sixth order we can have
another set of three equations which can be used to used to
determine all the relevant tree-amplitudes.
\begin{equation}
\langle \delta^2 ({\bf x}_1)  \delta^2 ({\bf x}_2)\delta^2 ({\bf
x}_3)\rangle = (8T_a + 8T_e)( \xi_{ab} \xi_{bc} + \xi_{ab}
\xi_{ac} + \xi_{ac} \xi_{bc})\bar{ \xi_2}^3 \end{equation}
\begin{equation}
\langle \delta ({\bf x}_1)  \delta^2 ({\bf x}_2)\delta^3 ({\bf
x}_3)\rangle = ((18 T_b + 12T_a + 6T_c) \xi_{ab} \xi_{bc} + (12T_a
+ 24T_b + 12T_e + 6T_d)\xi_{ab} \xi_{ac} + (6T_b + 12 T_a)\xi_{ac}
\xi_{ab})\bar{ \xi_2}^3 \end{equation} \begin{equation} \langle
\delta ({\bf x}_1) \delta ({\bf x}_2)\delta^4 ({\bf x}_3)\rangle =
(24T_a + 24T_e + 12T_b + 4T_d)( \xi_{ab} \xi_{bc} + \xi_{ab}
\xi_{ac})+ (4T_f + 24T_e + 36T_c + 48T_d + 24T_a + 120T_b)
\xi_{ac} \xi_{bc})\bar{ \xi_2}^3
\end{equation}
This method can be extended to higher-order correlation functions
and, combined with a judicious use of factorial correlators or
factorial moments described in Munshi et al. (1998b), it provides
the simplest way to determine the hierarchical amplitudes. It is however
to be noted that we have neglected the small correction factors
pointed out by Boschan et. al. (1994).

\subsection{Generalised Cumulant Correlators in the Bernardeau--Schaeffer Ansatz}
As explained earlier, although all hierarchical models (2) agree
on the basic tree structure underpinning the correlation hierarchy
describing the matter distribution, specific models differ from
each other in the assumptions they make regarding the amplitudes
associated with tree-topologies. One such model, which has been
studied in great detail, was proposed by Bernardeau \& Schaeffer
(1992). They assumed that each vertex (or node) in the tree
representation of correlation functions (2) carries an weight
which depends only on the order of the vertex and on nothing else.
Vertices of the same order appearing in different trees will carry
exactly same weight. The amplitude of the whole tree is
consequently always equal to the product of the weights assigned
to each constituent node. Even with this rather simple assumption
this model can predict many interesting features associated with
collapsed objects, some of which have already been tested
successfully against numerical simulation (Munshi et al. 1998b).
More sophisticated models have also been constructed by assuming a
specific form for the generating functions of these vertex
amplitudes. Such models can make further predictions about many
body statistics of collapsed objects.

Assuming that the tree amplitudes can be factorized in this way,
it is possible to decompose the cumulant correlators of various
orders in the  following way (Bernardeau 1992; Munshi et. al.
1998a): for the two-point function of arbitrary order
\begin{equation}
\langle \delta^p({\bf x}_a) \delta^q({\bf x}_b) \rangle =
[C_{p1}\xi_{ab}C_{q1} ] {\bar \xi}^{p + q -2};
\end{equation}
for three points,
\begin{equation}
\langle \delta^p({\bf x}_a) \delta^q({\bf x}_b) \delta^r({\bf
x}_c) \rangle = \big [ C_{q1} \xi_{ab} C_{p11} \xi_{ac} C_{r1} +
C_{p1} \xi_{ab} C_{q11} \xi_{bc} C_{r1} + C_{p1} \xi_{ac} C_{r11}
\xi_{bc} C_{p1} \big ] {\bar \xi}^{p + q +r -3};
\end{equation}
for four points
\begin{eqnarray}
\langle \delta^p({\bf x}_a) \delta^q({\bf x}_b) \delta^r({\bf
x}_c) \delta^s({\bf x}_d) \rangle & = & \big [ C_{q1} \xi_{ab}
C_{p111} \xi_{ad} C_{s1} \xi_{ac} C_{r1} + ({\rm
cyclic~permutations}) \nonumber\\ & & +
C_{q1}\xi_{ab}C_{p11}\xi_{ac}C_{r11} \xi_{cd} C_{s1} +
  ({\rm cyclic~permutation}) \big ]{\bar \xi}^{p + q +r + s -4};
\end{eqnarray}
and for five  points
\begin{eqnarray}
\langle \delta^p({\bf x}_a) \delta^q({\bf x}_b) \delta^r({\bf
x}_c) \delta^s({\bf x}_d) \delta^t({\bf x}_e) \rangle & = & \big [
C_{p1111}\xi_{ab}C_{q1}\xi_{ac}C_{r1}\xi_{ae}C_{t1} + ({\rm
cyclic~permutation}) \nonumber\\ & & +~
C_{p111}\xi_{ad}C_{q11}\xi_{bc}C_{r1}\xi_{ad}C_{t1}\xi_{ae}C_{s1}
 + ({\rm cyclic~permutation}) \nonumber\\
 & & +~C_{p1} \xi_{ab} C_{q11}
\xi_{bc} C_{r11} \xi_{cd} C_{s11} \xi_{de} C_{t1} + ({\rm
cyclic~permutation}) \big ] {\bar \xi}^{p + q +r + s +t -5}
\end{eqnarray}
The vertex functions $C_{p1 \dots 1}$ are simplest to deal with
when expressed in terms of their generating functions, e.g.
\begin{equation}
\mu_1(t) = \sum_{p=1}^{\infty} \frac{C_{p1} t^p}{p!},
\end{equation}
\begin{equation}
\mu_2(t) = \sum_{p=1}^{\infty} \frac { C_{p11} t^p }{p!},
\end{equation}
and so on. With these definitions it is possible to relate the
generating functions $\Psi^{p}(t_1, \dots, t_p)$ of the cumulant
correlators with generating functions of the vertex weights (see
Munshi et al. 1998a for more details):
\begin{equation}
\Psi^{(2)}(t_1,t_2) = \mu_1(-t_1) \xi_{ab} \mu_1(-t_2)
\end{equation}
\begin{equation}
\Psi^{(3)}(t_1,t_2,t_3) = \mu_1(-t_1) \xi_{ab} \mu_2(-t_2)\xi_{ac}
\mu_1(-t_3) + \dots {\rm (cyclic~permutations)}
\end{equation}
\begin{eqnarray}
\Psi^{(4)}(t_1,t_2,t_3, t_4) & = & \mu_1(-t_1) \xi_{ab}
\mu_3(-t_2)\xi_{ac} \mu_1(-t_3)\xi_{ad} \mu_1(-t_4) + \dots {\rm
(cyclic~permutations)}  \nonumber \\ & &
+ \mu_1(-t_1) \xi_{ab}
\mu_2(-t_2)\xi_{ac} \mu_2(-t_3)\xi_{ad} \mu_1(-t_4) + \dots {\rm
(cyclic~permutations)}
\end{eqnarray}
Specific forms for $\mu_n(t)$ have been obtained both by
diagrammatic techniques and direct analytical computation in
Munshi et al. (1998a). Direct evaluation of the multi-point
cumulant correlators provides us with a way to check such
predictions to arbitrary order. For example, according to the
model of Bernardeau \& Schaeffer (1992), the topological
amplitudes at higher order are always related with the lower order
amplitudes in a specific way: at every order this particular model
introduces a new ``star'' diagram, and all other diagrams at every
order can be expressed in terms of lower order amplitudes in a
simple fashion. For example, in this model, $R_a$ (4th order) is
generated by $Q  = Q_3$ (third):
\begin{equation}
R_a = Q^2,
\end{equation}
while $R_b$ is a ``new'' amplitude at fourth order, which cannot
be related to $Q$. At fifth order,
\begin{equation}
S_a = Q^3, S_b = QR_b \end{equation} but $S_c$ is not constrained
by $Q$ or $R_a$ or $R_b$. Likewise,
\begin{equation}
T_a = Q^4,\, T_b = Q^2R_b, \, T_c = R_b^2, \,T_d = QS_c, \, T_e =
Q^2R_b,
\end{equation}
with $T_f$ free at this level. Tests of these predictions will
become possible as larger simulation boxes become available. These
will undoubtedly lead us to better understanding of how this model
describes the nonlinear regime of gravitational clustering.

\subsection{Generalised Cumulant Correlators in  the Szapudi--Szalay Ansatz}
An analysis similar to the preceding one has been carried out by
Szapudi \& Szalay (1993a), using a different ansatz for the
tree-level amplitudes to that of Bernardeau \& Schaeffer (1992).
Szapudi \& Szalay  (1993b) subsequently also looked at the
predicted statistics of over-dense cells in this light. In their
analysis they assume that amplitudes associated with different
topologies but with same number of vertices are always equal. For
example this means that, at fourth order, all tree topologies will
have an amplitude $Q_4$ which is a weighted average of different
topologies of fourth order. Since there are 16 distinct
configurations of 4 points, 12 of which are ``snake'' and four of
which are ``star'', this means  $16Q_4 = 12R_a + 4R_b$. Similarly,
at fifth order, $125Q_5 = 60S_a +60S_b+5S_c$ and in general $Q_{N,\alpha} = Q_N$. The
corresponding formulae are:
\begin{equation}
\langle \delta^p({\bf x}_a) \delta^q({\bf x}_b) \rangle =
p^{p-1}q^{q-1}Q_{p+q}\xi_{ab} {\bar \xi}^{p + q -2}
\end{equation}
\begin{equation}
\langle \delta^p({\bf x}_a) \delta^q({\bf x}_b) \delta^r({\bf x}_c) \rangle
= p^{p-1}q^{q-1}r^{r-1}Q_{p+q+r}\big [ p \xi_{ab}\xi_{ac}  +
q\xi_{ab} \xi_{bc}  +  r\xi_{ac}  \xi_{bc}  \big ] {\bar \xi}^{p + q +r -3}
\end{equation}
\begin{eqnarray}
\langle \delta^p({\bf x}_a) \delta^q({\bf x}_b) \delta^r({\bf
x}_c) \delta^s({\bf x}_d) \rangle & = &
p^{p-1}q^{q-1}r^{r-1}s^{s-1}Q_{p+q+r+s}\big [ p^2 \xi_{ab}
\xi_{ac}  \xi_{ad}  + ({\rm cyclic~permutations})
 \nonumber \\
& & + ~ pr \xi_{ab}\xi_{ac}
 \xi_{cd}  + ({\rm cyclic~permutations}) \big ]{\bar \xi}^{p + q +r + s -4}
\end{eqnarray}
\begin{eqnarray}
\langle \delta^p({\bf x}_a) \delta^q({\bf x}_b) \delta^r({\bf
x}_c) \delta^s({\bf x}_d)
 \delta^t({\bf x}_e) \rangle & = &
p^{p-1}q^{q-1}r^{r-1}s^{s-1}t^{t-1}Q_{p+q+r+s+t}\big [
p^3\xi_{ab}\xi_{ac}\xi_{ad}\xi_{ae} + ({\rm
cyclic~permutations})\nonumber\\ & & +
~p^2q\xi_{ab}\xi_{bc}\xi_{ad}\xi_{ae}
 + ({\rm cyclic~permutations})   \nonumber \\
 & & + ~qrs \xi_{ab} \xi_{bc} \xi_{cd} \xi_{de}
+ ({\rm cyclic~permutations}) \big ] {\bar \xi}^{p + q +r + s +t -5}
\end{eqnarray}
These results again furnish a simple test of this model. One
simply needs to evaluate $Q_{p+\dots+t}$ for different values of
the particular $p$, $q \ldots, t$ and check if the parameter
$Q_{p+\dots+t}$ remains invariant when the individual $p$,
$q\ldots t$ are changed in such a way that $p+\dots+t$ remains
constant.

Using two-point cumulant corrrelators, Munshi \& Melott (1998)
have recently tested predictions of these particular versions of
the generic hierarchical ansatz against numerical simulations
up-to fourth order. More detail studies using three-point
factorial correlators up to sixth order will be presented
elsewhere.

\begin{figure}
\protect\centerline{
 \epsfysize = 3.5truein
 \epsfbox[20 420 587 714]
 {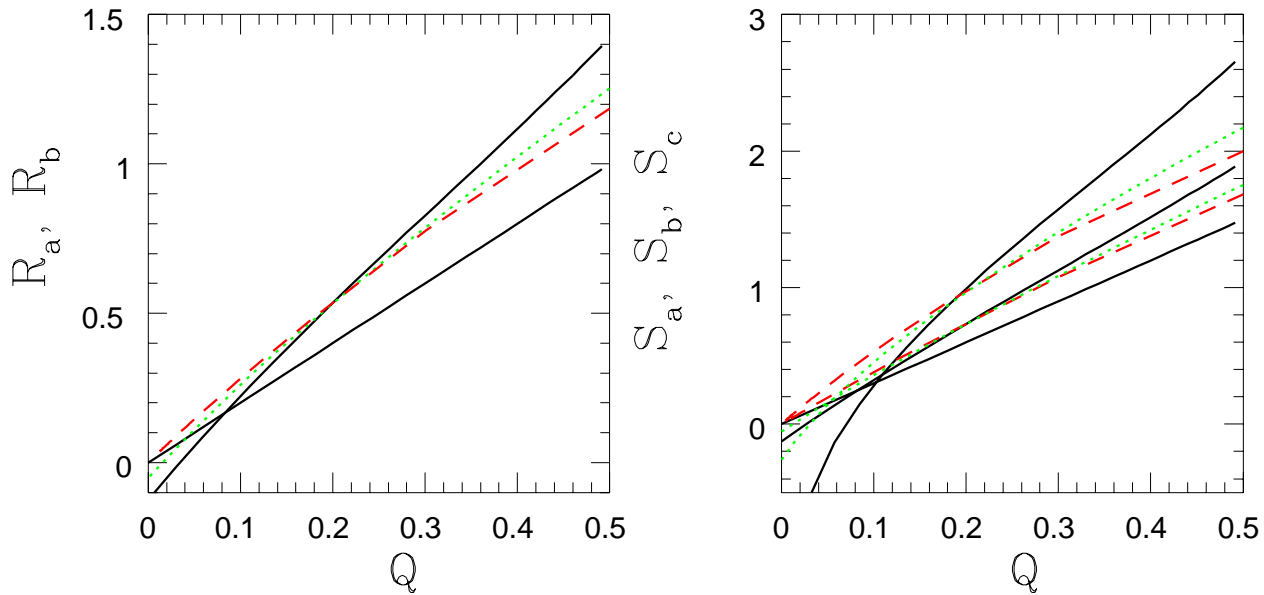} }
 \caption{Predictions for hierarchical
amplitudes of fourth order $R_a$, and $R_b$ and fifth order $S_a$,
$S_b$ and $S_c$ are plotted as a function lowest order
hierarchical amplitude $Q$. Solid lines are predictions from
hyper-extended perturbation theory (Scoccimarro \& Frieman 1998),
dotted lines represent extended perturbation theory (Colombi et
al. 1996) and dashed lines represent predictions from specific
hierarchical model of Bernardeau \& Schaeffer (1992). The bottom
most curve in each panel represents snake diagrams $R_a = Q^2$ and
$S_a = Q^3$ which are same for all these approximations are same.
The triplets of upper curves in left panel represent $R_b$ and in
right panel they represent $S_c$ and $S_b$ respectively from top
to bottom.}
\end{figure}

\subsection{Hierarchical Amplitudes from Extended and Hyper-extended Perturbation
Theory}
Extended perturbation theory and hyper-extended perturbation theory
have been suggested by Colombi et al. (1996) and Scoccimarro \&
Frieman (1998) respectively in attempts to obtain more accurate
analytic predictions of the $S_N$ parameters in the highly
non-linear regime. Such extensions however are, however, silent
concerning the specific contribution to $S_N$ from distinct
topologies. We can improve upon this situation, by combining
predictions of the hierarchical model of Bernardeau \& Schaeffer
(1992) for the highly non-linear regime with these predictions
using (hyper)extended perturbation techniques.

It turns out that this procedure always predicts that ``star''
topologies carry more weight than ``snake'' and ``hybrid''
topologies. We also find that the amplitudes of all topologies
increases when the fluctuations have more large-scale power. For
power spectra with less power on larger scales we find that all
topological amplitudes become approximately equal, although perturbation
theory is known to break down for such spectra. Existing
numerical results at fourth order ($R_a$ and $R_b$) roughly match
with these predictions (Munshi \& Melott 1998). However, more
tests are needed to check predictions at orders greater than four.

It may not be possible to extend tree-level perturbation
techniques straightforwardly by changing spectral index $n$ to
effective spectral index $n_{\rm eff}$ in order to compute
multi-point cumulant correlators. In principle, although these
quantities can be computed for arbitrary points in the
quasi-linear regime they will in general imply that hierarchical
amplitudes are depend on shape parameters while, in the highly
nonlinear regime hierarchical amplitudes are thought to be
independent of shape. At least at the level of three-point this
seems to have been already confirmed by some numerical experiments
(Scoccimarro et al. 1998). Furthermore, it has been pointed out
(Bernardeau 1996) that, in the quasilinear regime,  tree-level
perturbation theory for one--point cumulant of the smoothed
density field $S_3$, implies $S_3 = 3Q$ but for $C_{21}$ is not
equal to $2Q$ as one would have expected on the basis of the
hierarchical ansatz (which applies to the unsmoothed tree-level
perturbation theory). Clearly, one has to be careful about the
effect of smoothing on the cumulant correlators. When combining
the hierarchical ansatz with extended or hyperextended
perturbation theory we have to keep these points in mind. It is
nevertheless interesting to combine these two kinds of calculation
in this way, as there are otherwise no analytical predictions for
these quantities that can be tested against numerical simulations.

The results of this and the previous calculations are shown in
Figure 1.

\section{Quasi-Linear Cumulant Correlators From Projected Catalogs}

In the quasi-linear regime, two-point cumulant correlators have
already been studied in great detail using perturbation theory.
The extension of such studies to multiple points, as is required
for the generalisations we discuss in this paper, are difficult
owing to the complicated shape dependence of the hierarchical
amplitudes. For this Section, therefore, we shall focus on the
(original) two-point version of these statistics.

Estimates of the cumulant correlators were extracted from the APM
catalogue  by Szapudi \& Szalay (1997) and a related theoretical
computation of projected cumulant correlators was done by
Bernardeau et al. (1997). Let us now concentrate on the APM
results in order to illustrate the excellent match between theory
and observation.

The usual notation with which the two--point cumulant correlators
are expressed is in terms of $C_{pq}$, are defined as (Bernardeau
1995; Szapudi \& Szalay 1997)
\begin{equation}
\langle\delta({\bf x}_1)^p \delta({\bf x}_2)^q\rangle_c = C_{pq}
\langle \delta^2({\bf x}) \rangle \langle \delta({\bf x}_1)
\delta({\bf x}_2) \rangle^{p+q-2}.
\end{equation}
An alternative parameterisation, which we shall in fact use,  is in
terms of the quantities
\begin{equation}
Q_{pq} = \frac{C_{pq}}{p^{p-1} q^{q-1}}
\end{equation}
We assume the initial density contrast $\delta({\bf x})$ (3) to be
Gaussian which means that it can be decomposed into independent
Fourier modes $\delta({\bf k})$
\begin{equation}
\delta({\bf x}) = \int d^3{\bf k} \delta({\bf k}) \exp( i{\bf
k}.{\bf x}),
\end{equation}
which completely characterized by power spectra $P(k)$
\begin{equation}
\langle \delta({\bf k}) \delta({\bf k'})\rangle = \delta_D({\bf k
+ k'})P(k).
\end{equation}
In such a field, each Fourier mode is statistically independent of
the others.

By analogy, for projected catalogs, one define angular cumulant
correlators to be
\begin{equation}
\langle\sigma({\bf \gamma}_1) \sigma({\bf \gamma}_2)\rangle_c =
c_{pq} \langle \sigma^2({\bf \gamma }) \rangle \langle \sigma({\bf
\gamma}_1) \sigma({\bf \gamma}_2) \rangle^{p+q-2},
\end{equation}
where ${\bf \gamma}_1$ and ${\bf \gamma}_2$ are unit vectors
defining positions in the celestial sphere. A quantity $q$ can be
defined with respect to $c$ in the same way as analogy with $Q$ is
defined with respect to $C$ in equation (33). The quantity
$c_{pq}$ is the projected cumulant correlator and $\sigma({\bf
\gamma})$ is projected number density of objects:
\begin{equation}
\sigma({\bf \gamma}) = n_0 \int_0^{\infty} r^2 dr F(r) { \rho( r
\gamma) \over \rho_0},
\end{equation}
where $F(r)$ is the selection function of the catalog and $\rho_0$
the mean density of the universe, $n_0 F(r)$ is mean density of
observable objects at a radial distance $r$. The projected density
field on the sky will usually be smoothed with a smoothing window
$W_{\gamma_0}$, the result of which can be written as
\begin{equation}
\sigma_s({\bf \gamma}) = n_0 \int W_{\bf{\gamma}_0}({\bf \gamma}_1)d^2{\bf \gamma}_1
\int r^2 dr F(r) { \rho(r {\bf \gamma}) \over \rho_0}.
\end{equation}
The function $W(\bf \gamma_0)$ is unity if the angle between $\bf
\gamma$ and a given direction ${\bf \gamma_0}$ is less than
$\theta_0$ and zero otherwise. The projected one--point cumulants
\begin{equation}
s_p = \langle \sigma_s^p({\bf \gamma}) \rangle / \langle
\sigma_s^2({\bf \gamma})\rangle^{p-1} \end{equation}
 are normalised moments of the smoothed one-point projected probability
density functions, and coincide with the cumulant correlators in
the limit $ {\bf x}_1 \rightarrow {\bf x}_2$.

The most important assumption that we will be using in our
derivation of $c_{pq}$ in projection is the fact that both
smoothing angle and the separation angle are small compared to
unity. In general, $c_{pq}$ will contain terms which are of higher
order in $\langle\sigma({\bf \gamma}_1) \sigma({\bf \gamma}_2)
\rangle$, but we will be neglecting all such terms in our
calculations. Detailed perturbative calculations are needed to
take higher order loop corrections into account. In this paper we
however focus on directly comparing these theoretical predictions
against measurements from the APM (Szapudi \& Szalay 1997). One of
the advantage of using small angle approximation is the separation
of effects due to selection function and that of dynamical
contributions. Using these approximations, it is possible to show
that the projected cumulant correlators can be written:
\begin{equation}
c_{pq} = R_{p+q} \Theta_{pq};~~~~~~~ q_{pq} =
c_{pq}/q^{q-1}p^{p-1};
\end{equation}
where $\Theta_{pq}$ are the cumulant correlators derived from the
simplified dynamics of two-dimensional isotropic collapse in a
manner we shall shortly describe. We can define the generating
function $\beta(y_1, y_2)$ for $\Theta_{pq}$ as
\begin{equation}
\beta(y_1, y_2) = \sum_{p=1, q=1}^{\infty} \Theta_{pq} { y_1^p
y_2^q \over p! q!}.
\end{equation}
Since we are interested only in terms linear in $\langle\delta(\bf{\gamma}_1) \delta(\bf{\gamma}_2)\rangle$ we can write
$\beta(y_1, y_2) = \beta(y_1) \beta(y_2)$ (Bernardeau 95), where
\begin{equation}
\beta(y_1) = \sum_{p=1}^{\infty} \Theta_{p1} { y_1^p \over p!},
\end{equation}
which means that, in linear order, $\Theta_{pq} =
\Theta_{p1}\Theta_{q1}$. Bernardeau (1992) showed that the
generating function,
\begin{equation}
G(\tau) = \sum_{p=1}^{\infty} \nu_p \frac{(-\tau)^p}{p!},
\end{equation}
of the tree amplitudes $\nu_p$ that appear in a perturbative
expansion of the density distribution satisfy dynamical equations
governing collapse of spherical over density, with $\tau$ playing
the role of time (Bernardeau 1992; Munshi et al. 1994). It can be
shown that generating function for cumulant correlators $\beta(y)$
can be related to $G{(\tau)}$ by the following relation
(Bernardeau 1995):
\begin{equation}
\beta(y) = -y \frac{d G(\beta(y))}{d \tau};
\end{equation}
\begin{equation}
\tau =- y \frac{ d G(\tau)}{d \tau}.
\end{equation}
These relations hold only for an {\em unsmoothed} initially
Gaussian field. In the case of angular smoothing and power law
initial condition $P(k) = Ak^n$ one can write
\begin{equation}
\beta(y) = \tau(y)\left[1 + G(\tau(y))\right]^{-(n+2)/2}.
\end{equation}
It is easy to see that although the dynamics is determined by two
dimensional spherical collapse, the spectral index $n$ is still
that of three dimensional power spectrum $P(k)$.

Expanding $\beta(y)$ in Taylor series we obtain the dynamical
contribution to the cumulant correlators (Bernardeau et al. 1997;
Munshi \& Melott 1998)
\begin{equation}
\Theta_{21} = \frac{24}{7} - \frac{1}{2} ( n+2)
\end{equation}
\begin{equation}
\Theta_{31} = \frac{2946}{98} - \frac{195}{14}(n+2) +
\frac{3}{2}(n+2)^2.
\end{equation}

The factors $R_{p+q}$ in (40) depend on the spectral index and
specific form of selection function. They were determined by
Gaztanaga (1994)
\begin{equation}
R_p = {m_1^{p-2}(3)M_p ( 3p -(p-1)(n+3)) \over M_2^{p-1}(3-n)}
\end{equation}
where
\begin{equation}
M_p(a) = \int_0^{\infty} dr~r^{a-1} F^p(r).
\end{equation}
Using the specific form for the selection function $F(r) = K
r^{-0.5}\exp [-(r/D)^2]$ gives $R_3 = 1.19$, $R_4 = 1.52$, $R_5 =
2.00$, as computed by Gaztanaga (1994) and Bernardeau (1995). The
slope of the projected two-point correlation function used for
comparison is $\gamma = 1.7$.

\begin{figure}
\protect\centerline{ \epsfysize=3.5truein
 \epsfbox[20 146 587 714]
{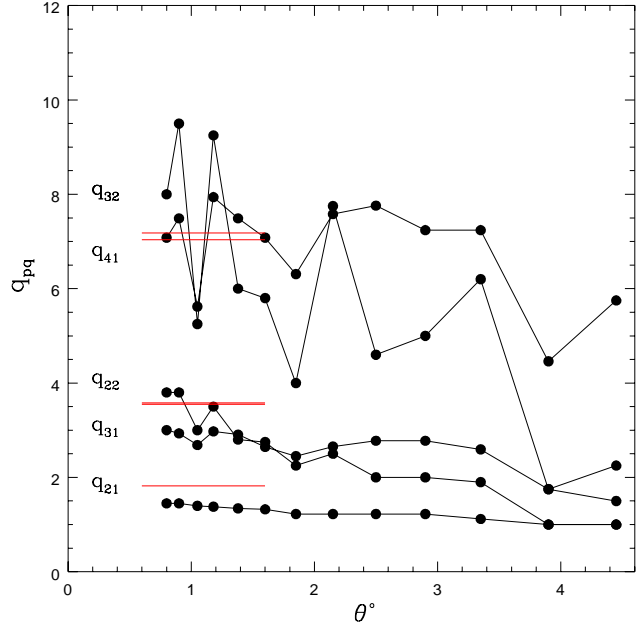}
}
\caption{Values of $q_{pq}$ measured
from the APM survey by Szapudi \& Szalay (1997) as a function of
separation angle $\theta^\circ$. The solid lines represent
tree-level perturbative predictions for small angles. }
\end{figure}

\section{Comparing Analytical Calculations with APM measurements}
Fully nonlinear calculations of $q_{pq}$ using factorial moments
were performed by Szapudi \& Szalay (1997), who used $0.23^\circ$
cells to construct density maps with a magnitude cuts of $b_J =
17$ to $20$. Measurements of $q_{pq}$ were presented up-to fifth
order. For these magnitude cuts, the angular separation of
$1^{\circ}$ corresponds approximately to $7h^{-1}$Mpc, a regime in
which where perturbative calculations should be valid. We have
replotted $q_{pq}$ measured from the APM survey by Szapudi \& Szalay
(1997) as a function of separation angle $\theta^\circ$. The
smoothing angle remains fixed for all separation angles.

The theoretical predictions discussed above are plotted as
straight lines and show reasonable agreement with measured values.
Our results are valid in the limit of large separations and in the
limit of small angles. These approximations need to be tested
with more detailed perturbative calculations and Monte-Carlo
simulations. It is clear that dominant contribution will come from
the tree level approximation used by us. Simultaneous use of these
two approximations may appear self contradictory, but it has been
shown by direct Monte-Carlo integration of $\xi_3(r_1, r_2, r_3)$
in real space that the value of $C_{21}$ converges to its value in
the  large separation limit,  even when the separation of the two
cells is only equal to their diameter (Bernardeau 1995). It has
also been shown (Gaztanaga \& Bernardeau 1997), that the
small-angle approximation is very good  for smoothing scales less
than $1^\circ$ and given the errors associated with measurements
from galaxy catalogs are large, it provides a very good
approximation even for separation as large as $5^\circ$. On the
other hand, it is expected that contributions coming from  terms of
higher order in $\langle\sigma({\bf \gamma}_1) \sigma({\bf
\gamma}_2) \rangle$ will increase the tree-level values computed
here. Finite sample volume corrections tend to reduce the computed
value of cumulant correlators and enhance the agreement with
tree-level perturbative calculations.

It will be very interesting to include effects of finite volume
corrections in measurement of cumulant correlators from galaxy
surveys and compare with perturbative calculations beyond tree
level. However, the important point we wish to stress here is that
projection effects make it difficult to {\em deduce} the values of
hierarchical amplitudes, such as $R_a$ and $R_b$, directly. This
means that a comparison with predictions made in real space and in
three dimensions with projected cumulant correlators is highly
problematic (e.g. Szapudi \& Szalay 1997).

It is also interesting to note that, although the smoothing length
involved in this measurement is in the quasi-linear regime, the
non-linear correction process adopted by Szapudi \& Szalay is
based on a hierarchical ansatz that applies in the highly
nonlinear regime. This nevertheless seems to work very well as the
values of $r_a$ and $r_b$ (the projected amplitude of two
different topologies) derived show a constant amplitude over all
angular separations after correction. However, there may be an
error in their paper. They have derived the relations $r_a =
q_{22}$ and $r_b = 3 q_{31}-2 q_{12}$ for large angular separation
$\theta$, as all other terms higher order in $\xi_l/\xi_s$ are
negligible in this limit. Taking a closer look in their figure we
can see that while the measured value of $q_{22}$ for large
angular separations in the lower panel is very close to unity the
measured value for $q_{31}$ in the same limit is close to $2.23$
(approximately 0.35 in logarithmic units). Using these values of
$q_{31}$ and $q_{21}$ we get $r_a = q_{22} \simeq 1$ and $r_b = 3
q_{31}-2q_{12} \simeq 3\times2.23 - 2\times 1 \simeq 4.69$.
However the measured $r_a$ remains close to $5.5$, and the curve
which they label $r_b$ remains close to unity.  This indicates
that they have probably inadvertantly interchanged $r_a$ and
$r_b$.

\section{Discussion}

We have shown how to generalise the concept of a cumulant
correlator to an arbitrary number of points, and have
deconstructed the resulting functions explicitly up to sixth
order in large separation limit. These results were obtained without making any
specific
ansatz for amplitudes associated with different tree-topologies.
These results, when combined with results of multi-point factorial
correlators derived in Munshi et al. (1998b), allow us to build a
direct determination of the hierarchical amplitudes. They can also
be used to test the dependence of these parameters on shape
parameters in the highly nonlinear regime.

Earlier studies by Szapudi \& Szalay (1997) showed that two-point
cumulant correlators can separate contributions from tree
topologies at fourth order. In this paper we have explicitly shown
that if we move from two-point to three-point cumulant correlators
it will be possible to separate all tree topologies contributing
to fifth and sixth order correlation functions. For higher orders
it is necessary to move to four-point cumulant correlators, and so
on.

We have also studied the predictions of more particular versions
of the hierarchical ansatz which generally assume some specific
form for the hierarchical amplitudes. In particular, we have
studied an extension of the ansatz given by Bernardeau \&
Schaeffer (1992) which assumes that, in the highly nonlinear
regime, a given tree amplitude can be constructed by multiplying
vertex amplitudes constituting the tree. We have also extended the
ansatz by Szapudi \& Szalay (1992) which is based on replacing
each amplitude of the different tree-topologies by an average over
different topologies of same order. With our analysis it will be
possible to test these results against numerical simulations when
simulations with larger dynamic range become available. This will
provide an unique way to study gravitational clustering in the
highly nonlinear regime.

While using our results of multi-point cumulant correlators
derived in highly nonlinear regime it should be realized that
although the decomposition is still valid, the amplitudes depend
strongly on shape factors. The shape dependence of lowest order
tree amplitude $Q_3$ has been studied extensively by Scoccimaro
et al. (1998). Similar studies of higher order amplitudes will
clarify weather these quantities do become independent of shape
factors and also more studies are needed to check how they depend
on initial power spectra.

We have shown that predictions made for two--point cumulant
correlators match very well when compared with measurement from
projected catalogs in the quasilinear regime. Our results in the
highly nonlinear regime will also be interesting when larger
three-dimensional galaxy catalogs are available. It will possible
to separate lower order tree topologies using our method and to
test the validity regime of different nonlinear approximations.

Hierarchical amplitudes are also related with the statistics of
collapsed objects, i.e. how the over-dense cells are distributed
against background matter distribution (Bernardeau \& Schaeffer
1992; Munshi et al. 1998a,b) so study of statistics of collapsed
objects provide another interesting way to determine the
hierarchical amplitudes. The reader is referred to Munshi et al.
(1998a,b) for further information about this work.

\section*{Acknowledgment}
DM acknowledges support from PPARC under the QMW Astronomy Rolling
Grant GR/K94133. ALM acknowledges the support of the NSF-EPSCoR
program. PC received a PPARC Advanced Fellowship during the period
when most of this work was performed.

%\end{document}

\bigskip
\bigskip
\bigskip

\begin{table*}
\begin{center}\rotate[l]{
\begin{tabular}{@{}|c|c|c|c|c|c|c|c|c|}
\multicolumn{9}{c}{Table 1. Multipoint Cumulant Correlators} \\ \\
\hline $\langle \delta^p \rangle$&$\langle \delta^p(x_a)
\delta^q(x_b)\rangle$&$ \langle \delta^p(x_a) \dots
\delta^r(x_c)\rangle$&$ \langle \delta^p(x_a) \dots
\delta^s(x_d)\rangle$&$ \langle \delta^p(x_a) \dots
\delta^t(x_e)\rangle$&$ \langle \delta^p(x_a) \dots
\delta^u(x_f)\rangle$&Amplitudes&Order&No. of Eq.\\ \hline \hline
$(p)$&$(p,q)$&$(p,q,r)$&$(p,q,r,s)$&$(p,q,r,s,t)$&$(p,q,r,s,t,u)$&&&\\
\hline \hline $(1)$&$$&$$&$$&$$&&&p+q+ \dots = 1&1\\ \hline
(2)&(1,1)&&&&&&p+q+ \dots = 2&2 \\ \hline
(3)&(1,2)&(1,1,1)&&&&$Q$&  p+q+ \dots = 3&3\\ \hline
(4)&(1,3),(2,2)&(1,1,2)&(1,1,1,1)&&&$R_a, R_b$&  p+q+ \dots =
4&5\\ \hline
(5)&(2,3),(1,4)&(1,1,2),(1,1,3)&(1,1,1,2)&(1,1,1,1,1)&&$S_a, S_b,
S_c$&  p+q+ \dots = 5&7\\ \hline
(6)&(2,4),(3,3),(1,5)&(1,1,4),(1,2,3),(2,2,2)&(1,1,1,3),(1,1,2,2)&(1,1,1,1,2)&
(1,1,1,1,1,1)&$T_a, T_b, T_c, T_d, T_e$& p+q+ \dots = 6&11 \\
\hline
\\
\\
\\
%\multicolumn{9}{c}{Multipoint cumulant correlators can be used to determine the hierarchical amplitudes associated with gravitational clustering in highly nonlinear regime  when they are believed to be} \\
%\multicolumn{9}{c}{independent of shape factors. It is clear that for a given order, multi-point cumulant correlators always depend on
%same number of hierarchical amplitudes. However with every additional new point
%the complication} \\
%\multicolumn{9}{c}{associated with evaluation of multi-point cumulant correlators
%increases considerably.} \\
%\multicolumn{9}{c}{ One point cumulants or $S_N$ parameters can not be
%used to decompose different amplitudes beyond third order. } \\
%\multicolumn{9}{c}{Two point cumulant
%correlators can actually be used to determine amplitudes associated with
%topological diagrams of fourth order, i.e. $R_a$ and $R_b$. } \\
%\multicolumn{9}{c}{Similarly if we
%move to three-point cumulant correlators we can determine all topological
%amplitudes associated with fifth order and sixth order diagrams. } \\
%\multicolumn{9}{c}{In fifth
%order the number of degenerate multi-point cumulant correlators are
%actually more than the number of topologies.} \\
%\multicolumn{9}{c}{ Additional equations provide
%valuable constrain equations which check the validity of such hierarchical
%ansatz. } \\ \\
\end{tabular}
}
\end{center}
\end{table*}

\bigskip
\bigskip
\bigskip

\newpage

\begin{thebibliography}{}
\bibitem{BaSa} Balian R., Schaeffer R., 1989, A\& A, 220, 1
\bibitem{B92} Bernardeau F., 1992, ApJ, 392, 1
\bibitem{} Bernardeau F., 1994, ApJ, 433, 1
\bibitem{} Bernardeau F., 1995, A\& A, 301, 309
\bibitem{} Bernardeau F., Schaeffer R., 1992, A\& A, 255, 1
\bibitem{} Bernardeau F., van Waerbeke L., Mellier Y., 1997, A\& A, 322, 1
\bibitem{} Boschan P., Szapudi I., Szalay A.S., 1994, ApJS, 93, 65
\bibitem{} Baugh C.M., Efstathiou G., 1993, MNRAS, 265, 145
\bibitem{} Baugh C.M., Gaztanaga E., 1996, MNRAS, 280, L37
%\bibitem{} Bouchet F.R., Strauss M.A., Davis M., Fisher K.B., Yahil A.,
%Huchra J.P.,
%1993, ApJ, 417
\bibitem{} Colombi S., Bouchet F.R., Hernquist L., 1996, ApJ, 465, 14
\bibitem{} Davis M., Peebles P.J.E., 1977, ApJS, 34, 425
\bibitem{} Fry J.N., 1984, ApJ, 279, 499
\bibitem{} Gaztanaga E., 1994, MNRAS, 268, 913
\bibitem{} Hamilton A.J.S, 1988, ApJ, 332, 67
\bibitem{} Juszkiewicz R., Bouchet F.R.,  Colombi S., 1993, ApJ, 412, L9
\bibitem{} Kauffmann G.A.M., Melott A.L. 1992, ApJ 393, 415
\bibitem{} Limber D.N., 1954, ApJ, 119, 665
\bibitem{} Maddox S.J., Efstathiou G., Sutherland W.J., 1996,
MNRAS, 283, 651
\bibitem{} Maddox S.J., Sutherland W.J., Efstathiou
G., 1990, MNRAS, 246, 433
\bibitem{} Maddox S.J., Sutherland W.J., Efstathiou G., Loveday
J., 1990, MNRAS, 243, 692
\bibitem{} Meiksin A., Szapudi I., Szalay A., 1992, ApJ, 394, 87
\bibitem{} Munshi D., Sahni V., Starobinsky A.A., 1994, ApJ, 436, 517
\bibitem{} Munshi D., Bernardeau F., Melott A.L., Schaeffer R.,
1999, MNRAS, in press (astro-ph/9707009)
\bibitem{}  Munshi D., Melott A.L, 1998, preprint/astro-ph/9801011
\bibitem{} Munshi D., Coles P., Melott A.L., 1998a, (submitted to MNRAS)
\bibitem{} Munshi D., Coles P., Melott A.L., 1998b, (submitted to MNRAS)
\bibitem{} Peebles, P.J.E., 1980, {\em The Large Scale Structure of the
Universe}. Princeton University Press, Princeton
\bibitem{} Scoccimarro R., Colombi S., Fry J.N., Frieman J.A., Hivon E.,
Melott A.L., 1998, ApJ, 496, 586
\bibitem{} Scoccimarro R., Frieman J., 1998, preprint,
astro-ph/9811184
\bibitem{} Szapudi I., Colombi S., 1996, ApJ, 470, 131
\bibitem{} Szapudi I., Dalton G., Efstathiou G.P., Szalay A.,
1995, ApJ, 444,520
\bibitem{} Szapudi I., Szalay A.S., 1993a, ApJ, 408, 43
\bibitem{} Szapudi I., Szalay A.S., 1993b, ApJ, 414, 414
\bibitem{} Szapudi I., Szalay A.S., 1997, ApJ, 481, L1
\bibitem{} Szapudi I., Szalay A.S., Boschan P., 1992, ApJ, 390, 350
\end{thebibliography}
\end{document}